\documentclass[aps,prd,twocolumn,showpacs,preprintnumbers,amsmath,amssymb,longbibliography]{revtex4-2}
\usepackage{hyperref}
\usepackage{bm,bbm}
\usepackage{graphicx}
\usepackage{braket}
\usepackage{mathtools}
\usepackage{mathrsfs}
\usepackage{comment}
\usepackage{autobreak}

\usepackage[utf8]{inputenc}

\usepackage{color}
\usepackage[dvipsnames]{xcolor}

\usepackage{tikz}
\usetikzlibrary{arrows,arrows.meta,intersections, calc,positioning,decorations.pathreplacing,decorations.pathmorphing,shapes}
\usetikzlibrary{patterns}
\usetikzlibrary{decorations.markings}
\usetikzlibrary{knots}

\begin{document}

\title{AdS gravastar and its signatures from dual conformal field theory}

\author{Heng-Yu Chen,$^{a}$ Yasuaki Hikida$^b$ and Yasutaka Koga$^b$}

\affiliation{$^a$Department of Physics, National Taiwan University, Taipei 10617, Taiwan}
\affiliation{$^b$Department of Information and Computer Science, Osaka Institute of Technology, Kitayama, Hirakata,  Osaka 573-0196, Japan}


\begin{abstract}

Quantum gravity effects are expected to resolve the black hole singularity and the effects may deform the region near but outside the horizon. Applying AdS/CFT correspondence, we see their signatures from the viewpoint of dual conformal field theory. As a regularized geometry, we consider AdS gravastar constructed by gluing AdS-Schwarzschild 
and de Sitter spacetime. The retarded Green functions of dual conformal field theory have bulk-cone singularities associated with null trajectories in the bulk and we obtain the singularities specific to a horizon-less geometry. We also observe 
echoes coming from  waves reflected behind the photon sphere. The existence of echoes implies the modification of geometry inside the photon sphere.

\end{abstract}

\maketitle


\section{Introduction}

Understanding quantum gravity effects is an important open problem in theoretical physics. Quantum gravity effects are expected to resolve black hole and cosmological singularities, but it is known to be difficult to examine these problems directly within gravity theory. The holographic principle offers a way to study gravity through a dual field theory \cite{Susskind:1994vu}. The dual field theory typically does not include gravity and is therefore more tractable than the gravity theory itself. The AdS/CFT correspondence provides a concrete realization of the holographic principle \cite{Maldacena:1997re,Gubser:1998bc,Witten:1998qj}, where a gravity theory on an asymptotically anti-de Sitter (AdS) spacetime is conjectured to be described by a conformal field theory (CFT). Careful studies of the dual CFT are expected to reveal quantum effects of the gravity theory.

Let us suppose a gravity system that we would like to investigate. To apply the AdS/CFT correspondence, we first realize this gravity system in an asymptotically AdS spacetime, which may be regarded as a toy model of the original system. It is therefore important to show that this toy model captures the essential qualitative features of the original system. Once this is established, the properties of the gravity system can be studied through the dual CFT, which may have a concrete definition or even an experimental realization.

It is expected that quantum gravity effects resolve the black hole singularity, and in some proposals like the ones with firewall \cite{Almheiri:2012rt} and fuzzballs \cite{Lunin:2001jy,Lunin:2002qf}, the region near but outside the horizon is modified. 
AdS-Schwarzschild black hole has a photon sphere and it is difficult to examine the near horizon region inside the photon sphere from an outside observer.
As a model of regularized geometry, we consider a spacetime with the photon sphere but without the black hole horizon. This kind of geometry may be provided by exotic compact objects (ECOs), such as boson star, wormhole etc., see, e.g., \cite{Cardoso:2019rvt}. 
Among them, we consider a gravastar solution as an example of asymptotic anti-de Sitter spacetime, see, e.g., \cite{Mazur:2001fv,Visser:2003ge,Pani:2009ss,Cardoso:2014sna} for asymptotic flat gravastar. AdS gravastar is constructed by replacing the region inside the photon sphere by de Sitter (dS) spacetime. We observe at least two types of signatures of AdS gravastar from the dual CFT, 
which can be regarded as typical signatures when black hole horizon in the bulk is resolved in some way. 
This work is part of a series of studies, and a more detailed and extended analysis can be referred to \cite{Chen:2025cee,Chen:2025jbf}.

Considering a CFT on $\mathbb{R}_t \times S^{d-1}$, where $\mathbb{R}_t$ and $S^{d-1}$ denote the time direction and $(d-1)$-dimensional sphere, respectively.
We examine the retarded Green function of a scalar operator,
\begin{align} \label{eq:defretarded}
    G_\text{R} (t , \theta) = i H(t) \langle [\mathcal{O} (t , \theta) , \mathcal{O} (0,0)] \rangle ,
\end{align}
where $H(t)$ is the Heaviside step function and the metric of $S^{d-1}$ is 
\begin{align}
\label{eq:metricS}
    d\Omega_{d-1}^2 = d \theta^2 + \sin ^2 \theta d \Omega_{d-2}^2 . 
\end{align}
We evaluate the retarded Green function from the propagation of a bulk scalar field following \cite{Son:2002sd}. There are light-cone singularities when the two insertion points of operators become light-like separated in the boundary geometry. Along with the usual singularities, the retarded Green function develops singularities when the bulk geodesic of the scalar particle becomes null. These are called bulk-cone singularities. See \cite{Hubeny:2006yu,Maldacena:2015iua,Hashimoto:2018okj,Hashimoto:2019jmw,Dodelson:2020lal,Kinoshita:2023hgc,Terashima:2023mcr,Hashimoto:2023buz,Dodelson:2023nnr,Caron-Huot:2025she,Chen:2025cee,Chen:2025jbf} for AdS-Schwarzschild black hole, where the null geodesic goes around the photon sphere. We find bulk-cone singularities associated with null geodesics traveling into the region inside the photon sphere for AdS gravastar. It is also known that echoes are typical signatures of gravitating objects with photon sphere but without horizon, see, e.g., \cite{Cardoso:2016rao,Cardoso:2016oxy,Oshita:2018fqu,Oshita:2020dox,Terashima:2025tct} for asymptotic flat case.
We observe echoes arising from the wave function of a bulk scalar field localized inside the photon sphere
for AdS gravastar.

The organization of this note is as follows.
In the next section, we review the computation of the retarded Green function from the wave function of a bulk scalar field. We employ the WKB approximation to solve the bulk wave equation.
In section \ref{sec:coordinate}, we transform the expression obtained in the momentum basis in section \ref{sec:mometum} into the coordinate basis. In this representation, bulk-cone singularities can be clearly identified.
For the AdS–Schwarzschild black hole, our result agrees with that obtained in \cite{Dodelson:2023nnr} up to an overall normalization factor. However, their analysis crucially relies on the presence of a black hole horizon and therefore cannot be applied to horizon-less geometries.
In contrast, our method applies to more general geometries, including AdS gravastars, and this technical development constitutes one of the main results of this work.
In section \ref{sec:gravastar}, we apply our method to the AdS gravastar. In addition to bulk-cone singularities specific to horizon-less objects, we observe echoes associated with wave functions localized inside the photon sphere.
In section \ref{sec:numerical}, we compare our analytic results with numerical computations and confirm their consistency.
Section \ref{sec:conclusion} is devoted to the conclusion and discussion. The appendix collects details of the WKB analysis of the wave equations.

In \cite{Chen:2025cee}, we constructed a four-dimensional AdS gravastar and numerically computed the retarded Green function for this geometry.
In the current work, we develop a semi-analytic technique to evaluate the retarded Green function and apply it to AdS gravastars in general dimensions.
We also present numerical results for the retarded Green function.
Further details of the analysis, as well as its extension to the AdS wormhole case, can be found in \cite{Chen:2025jbf}.

\section{Retarded Green functions}
\label{sec:mometum}

For $(d+1)$-dimensional bulk geometry, we use the metric of the form
\begin{align} \label{eq:metric}
    ds^2 = - f(r) dt^2 + f (r)^{-1} dr^2 + r^2 d \Omega_{d-1}^2 ,
\end{align}
where the metric of $S^{d-1}$ is \eqref{eq:metricS}. The metric function is
\begin{align} \label{eq:AdSS}
        f(r) = r^2 + 1 - \frac{\mu}{r^{d-2}}
\end{align}
for AdS-Schwarzschild black hole.
We consider the mode expansion of the bulk scalar field,
\begin{align}
 \phi (t , \Omega_{d -1} ,r) = e^{- i \omega t} Y_{\ell \vec m} (\Omega_{d-1}) r^{-\frac{d-1}{2}} \psi_{\omega \ell} (r) ,
\end{align}
where the spherical harmonics on $S^{d-1}$ are denoted by $Y_{\ell \vec m}(\Omega_{d-1})$.
The wave equation for the radial part can be put into the canonical form
\begin{align} \label{eq:wave}
    ( \partial_z^2 +  \kappa^2 (z) ) \psi (z) = 0 , \quad 
    \kappa (z) = \sqrt{\omega^2 -  V (z)} .
\end{align}
The new radial coordinate $z$ is introduced via
\begin{align}
    dz = - \frac{dr}{f(r)} .
\end{align}
For large $\ell$, the potential can be approximated by
\begin{align} \label{eq:potential}
 V(z) =  \left [\ell^2 + \left (\nu^2 - \frac{1}{4} \right) r^2 \right] \frac{f(r)}{r^2} .
\end{align}
Here $\nu = \sqrt{(d/2)^2 + m^2}$ with $m$ as the mass of the bulk scalar field and $\Delta = d/2 + \nu$ is the conformal dimension of the dual scalar operator.
Note that we include the term proportional to $r^2$ in the potential \eqref{eq:potential}, which is usually neglected, see, e.g., \cite{Dodelson:2023nnr}. 
The extra term will play an important role in later analysis around \eqref{eq:saddlept}.
In terms of $z$, the AdS boundary is located at $z \to 0$, where the scalar field behaves as
 \begin{align} \label{eq:waveSon}
     \psi (z) \sim \mathcal{A} (\omega , \ell) z^{\frac12 - \nu} + \mathcal{B} (\omega , \ell) z^{\frac12 + \nu} .
 \end{align}
From the asymptotic behavior, the retarded Green function can be obtained as \cite{Son:2002sd}
 \begin{align} \label{eq:GRSon}
     G_R(\omega , \ell) = \frac{\mathcal{B} (\omega , \ell)}{\mathcal{A} (\omega , \ell)}  .
 \end{align}

\begin{figure}
\centering
\includegraphics[height=4cm]{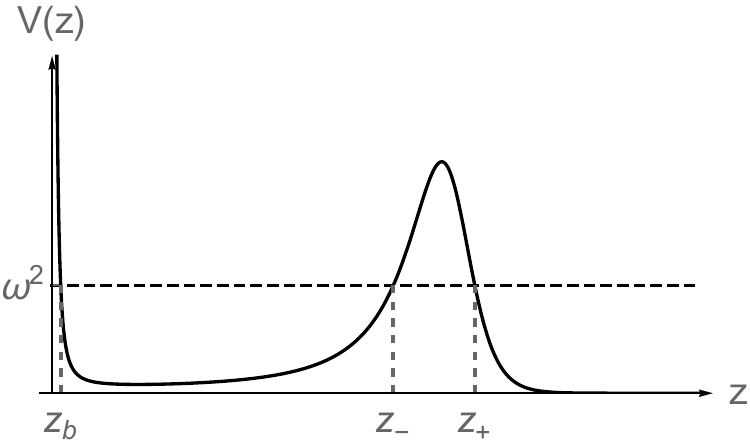}
\caption{The potential for AdS-Schwarzschild black hole}
\label{fig:Vsch}
\end{figure}

We first examine the wave equation \eqref{eq:wave} for AdS-Schwarzschild black hole. The potential is plotted in Fig.\,\ref{fig:Vsch}.
The black hole horizon is located at $r= r_h$ satisfying $f(r_h) = 0$ or at $z \to \infty$.
We assign the ingoing boundary condition at the horizon as
$\psi (z) \sim e^{i \omega z}$. 
For large $\ell$, the maximal of potential is realized by
\begin{align} \label{eq:Vmax}
\ell^{-2} V (r_c) \simeq  1 + \left( 1 - \frac{2}{d}\right) \left( \frac{2}{d \mu}\right)^{\frac{2}{d-2}}   ,  
\end{align}
where $r_c$ is the position of photon sphere,
\begin{align} \label{eq:rc}
     r_c = \left(  \frac{d \mu}{2} \right)^{\frac{1}{d-2}} .
\end{align}

We solve the wave equation \eqref{eq:wave} in the WKB approximation.
We consider the case where the equation $\kappa (z)^2 = 0$ has three zeros, see Fig.\,\ref{fig:Vsch}. Here $\kappa(z)$ was introduced in \eqref{eq:wave} and the three zeros are denoted as $ z_b , z_\pm$ $(z_b < z_- < z_+ )$. The potential diverges near the boundary $r \to \infty$ as $V(r) \propto r^2$, which leads to $\kappa(z)^2 = 0$ at $z = z_b $ with $z_b < 1/\ell$.
We start with the wave function  for $z_+ < z  < \infty$ satisfying the ingoing boundary condition at $z \to \infty$. Next we continue the wave function to the region with $0 \simeq z_b < z < z_-$. We can then read off the retarded Green function as in \eqref{eq:GRSon} from the asymptotic behavior of the wave function \eqref{eq:waveSon}.
The explicit form of the retarded Green function is given in \eqref{eq:retardedAdSG} and it is convenient to expand it in the form of
\begin{align} \label{eq:GRte}
G_R (\omega ,\ell) = \frac{\Gamma (- \nu)}{\Gamma(\nu)} \left( \frac{\omega^2 - \ell^2}{4}\right)^\nu \sum_{n=0}^\infty a_n e^{2 i n S(0,z_-)}\,. 
\end{align}
Here we have introduced the WKB phase,
\begin{align}
S(0,z_-)  =  \int_{0}^{z_-} \kappa (z') d z' ,
\end{align}
and assumed that $\text{Im} \, S(0, z_-) > 0$. 
The coefficients $a_n$ are 
\begin{align}  \label{eq:an}
a_0 =  e^{-i\pi \nu} , \quad a_{n} = - 2 i  (-1)^n e^{ -i n \pi \nu}  \sin ( \pi \nu ) 
\end{align}
with $n=1,2,3,\ldots$.
As shown below, each term with 
$n \geq 2$ 
corresponds to null geodesics that bounce off the AdS boundary 
 $n-1$ times.
For the moment, we neglect the terms suppressed by the factor $e^{- 2 S(z_- ,z_+)}$, where we set
\begin{align}
S(z_-,z_+)  =  \int_{z_-}^{z_+}  q (z') d z' 
\end{align}
with
\begin{align}
 q (z) = \sqrt{ V (z) - \omega^2 } .
\end{align}

\section{Bulk-cone singularities}
\label{sec:coordinate}

In order to observe the bulk-cone singularities, we perform the following Fourier transformation:
\begin{align} \label{eq:Fourier}
\begin{aligned}
G_R (t , \theta) = \frac{1}{2 \pi}  \sum_{\ell =0}^\infty \int_{- \infty + i \delta}^{\infty + i \delta} d \omega e^{-i\omega t} \\ \times 
G_R(\omega , \ell) \frac{\ell + \alpha}{\alpha} C_{\ell}^{(\alpha)}(\cos \theta) ,
\end{aligned}
\end{align}
where $C_{\ell}^{(\alpha)}(\cos \theta)$ are the Gegenbauer polynomials and $\alpha = (d-2)/2$.
Following the analysis in \cite{Dodelson:2023nnr}, the summand over $\ell$ is realized by an integration over a variable denoted as $k$ 
and the series expansion of Gegenbauer polynomials is applied. 
For $0 < \theta < \pi$, we can rewrite it as
\begin{align} \label{eq:sumj}
\begin{aligned}
G_R (t , \theta) = \sum_{j = 0}^\infty & \Big[g_R (t , |\theta| + 2 \pi j)  \\
& \quad + (-1)^{2 \alpha} g_R (t , 2\pi - |\theta| + 2 \pi j)\Big]
\end{aligned}
\end{align}
with
\begin{align} 
\begin{aligned}\label{eq:gE2}
    &g_R (t , \theta)  = \sum_{n=0}^\infty g_R^{(n)} (t , \theta) , \\
    &g_R^{(n)} (t , \theta) = \tilde a_n    \int_{\infty + i \delta}^{- \infty + i \delta} d \omega \int _{- \infty + i \delta '}^{\infty + i \delta '} d k  \\
& \quad \times  e^{- i \omega t + i k \theta} k^{\alpha}
 (\omega^2 -k^2)^\nu   e^{2 i n S (0,z_-)}  .
\end{aligned}
\end{align}
Here we have defined
\begin{align}
    \tilde a_n =  \frac{\Gamma (- \nu)}{\Gamma(\nu)} \frac{e^{- i\frac{\pi}{2 } \alpha + i \pi \alpha \lfloor \frac{\theta}{\pi} \rfloor}}{2^{\alpha +1 + 2 \nu} \pi |\sin \theta|^\alpha \Gamma(1+\alpha)}a_n .
\end{align}
The term with $n=0$ reproduces the light-cone singularity, see \cite{Dodelson:2023nnr}.
We thus focus on the terms with $n > 0$. 
We remark here that the integration over $k$ is performed over just above the real axis. In \cite{Dodelson:2023nnr}, the authors reduced the integral over $k$ as a sum over residues. However, this can be justified only when the contribution from large imaginary $k$ region can be neglected. This is the case for AdS-Schwarzschild black hole and is related to the existence of horizon. For generic geometries without horizon, the contribution from large imaginary $k$ region cannot be neglected.

We apply the saddle point approximation for the integrals in \eqref{eq:gE2}, where the exponent is 
\begin{align}
\Phi (\omega ,k)= - \omega t + k \theta + 2 n S(0,z_-) .
\end{align}
The saddle points of $\Phi(\omega ,k)$ can be obtained by $(\omega ,k) = (\omega_* , k_*)$ satisfying
\begin{align} \label{eq:saddlept}
\left.
\frac{\partial \Phi (\omega , k)}{\partial \omega}
\right|_{(\omega ,k) = (\omega_*, k_*)} = 
\left.
\frac{\partial \Phi (\omega , k)}{\partial k}
\right|_{(\omega ,k) = (\omega_*, k_*)} = 0 .
\end{align}
If we neglect the term proportional to $r^2$ in \eqref{eq:potential}, then the saddle points are degenerated and pinched singularities have to be treated as in \cite{Dodelson:2023nnr}.
The derivatives of the function $S(0,z_-)$ with respect to $\omega ,k$ can be approximated as (see \cite{Dodelson:2020lal,Chen:2025jbf})
\begin{align} 
\label{eq:TTnull}
\begin{aligned}
    &2  \frac{\partial S (0,z_-)}{\partial \omega}  \simeq T (u_*) - \frac{2 \bar \varrho u_*}{(u_*^2 - 1) k_*} , \\
    & 2 \frac{\partial  S (0,z_-)}{\partial k} \simeq - \Theta (u_*) + \frac{2 \bar \varrho}{(u_*^2 - 1) k_*} ,
    \end{aligned}
\end{align}
where we set $u_* = \omega_* / k_*$ and $\bar \varrho = \sqrt{\nu^2 - 1/4}$. 
Moreover, we have introduced
\begin{align} \label{eq:TTheta}
\begin{aligned}
&T(u) = 2 u \int_{r_*}^\infty \frac{dr}{f(r)} \frac{1}{\sqrt{u^2 - f(r) r^{-2} }} , \\
&\Theta(u) = 2  \int_{r_*}^\infty \frac{dr}{r^2} \frac{1}{\sqrt{u^2 - f(r) r^{-2}}} ,
\end{aligned}
\end{align}
which are the arrival time and angle of null geodesic, respectively.
Equation \eqref{eq:TTnull} can be derived in the following way. We first use the fact that $2  \frac{\partial S (0,z_-)}{\partial \omega}$ and $  2 \frac{\partial  S (0,z_-)}{\partial k}$ reduce to $T(u)$ and $-\Theta (u)$ in \eqref{eq:TTheta} for small $r$. We then approximate the region with large $r$ by pure AdS, which lead to the extra terms. For the argument, we assumed that $u_*^2 - 1 = \mathcal{O}((k_*)^0)$. 
This is expected to hold for generic geometries, but an important exception is the pure AdS case with $u_*^2 - 1 = \mathcal{O}((k_*)^{-1})$, see, e.g., \cite{Hubeny:2006yu}.

Applying the saddle point approximation, the term with $n > 0$ in \eqref{eq:gE2} can be written as
\begin{align} \label{eq:SPAm}
\begin{aligned}
g_R^{(n)} (t , \theta) &\simeq \tilde a_n k_*^\alpha ((\omega_*)^2 - (k_*)^2 )^\nu \frac{2 \pi ie^{ i \Phi (\omega_* , k_*) } }{\sqrt{\det H}}
 
 \end{aligned}
\end{align}
with
\begin{align} \label{eq:H}
H = \left. 
\begin{pmatrix}
\frac{\partial ^2 \Phi(\omega,k)}{\partial \omega \partial \omega} & \frac{\partial^2 \Phi(\omega,k)}{\partial \omega \partial k} \\ 
\frac{\partial ^2 \Phi(\omega,k)}{\partial k \partial \omega} & \frac{\partial ^2 \Phi(\omega,k)}{\partial k \partial k}
\end{pmatrix}
\right|_{(\omega,k) = (\omega_* , k_*)} .
\end{align}
We can show that
\begin{align}
 \Phi (\omega_* , k_*) = 0 
\end{align}
up to the order $\mathcal{O}(k_*^{-1})$ corrections.
Using \eqref{eq:TTnull} with \eqref{eq:TTheta}, we obtain the following
\begin{align} \label{eq:gRn}
    g_R^{(n)} (t, \theta ) & \simeq \tilde a_n\frac{(2 \bar \varrho n u_*)^{\alpha + 2 \nu + 3/2}}{(u_*^2 - 1)^{\alpha + \nu + 3/2}} \frac{2 \pi i}{\sqrt{ 2 \bar \varrho n^2 \frac{d}{du_*}T(u_*) }}\nonumber  \\
& \quad  \times  \frac{1}{( n T (u_*) - t)^{\alpha + 2 \nu + 3/2}} .
\end{align}
The bulk-cone singularities are located at  $t = n T (u_*)$, where $u_*$ is obtained by solving $\theta = n \Theta (u_*)$.
The corresponding null geodesic bounces off the AdS boundary $n-1$ times.
More detailed derivation will be presented in \cite{Chen:2025jbf}.
For $n=1$, the expression coincides with the one in \cite{Dodelson:2023nnr} up to an overall factor. 
However, we should emphasize that our arguments hold for generic geometries. 
In the case of pure AdS, the approximation in \eqref{eq:TTnull} cannot be applied and the light-cone and bulk-cone singularities coincide with each other.

\section{AdS gravastar and echoes}
\label{sec:gravastar}

We would like to apply the above analysis to the case of AdS gravastar. 
See \cite{Chen:2025cee} for detailed construction of four-dimensional gravastar.
We first consider the AdS-Schwarzschild black hole with the horizon at $r = r_h$ and the photon sphere at $r = r_c$ in \eqref{eq:rc}. We put a shell at $r =r_0$ with
\begin{align}
    r_h < r_0 < r_c 
\end{align}
and replace the core region $r < r_0$ by dS spacetime. We set the surface stress-energy tensor on the shell as
\begin{align}
\mathcal{S}=\sigma d\tau^2+\bar P r_0^2d\Omega^2_{d-1},
\end{align}
where $\sigma$ and $\bar P$ are the surface energy density and the surface tension, respectively. 
For simplicity, we set $\sigma = 0$.
We require that the induced metric is continuous. 
We further assign the junction condition
\begin{align} \label{eq:cond_energy}
    [[\chi]]-[[\text{tr}\, \chi]]h=-8\pi \mathcal{S} .
\end{align}
Here $\chi$ is the extrinsic curvature on the time-like hypersurface and the bracket $[[A]]=A_+-A_-$ denotes the jump of a quantity $A$ across the shell.
The metric of AdS gravastar can be put into the form  \eqref{eq:metric}, where the metric function $f(r)$ is
\begin{align}
\label{eq:gravastar-f}
    f(r)=\left\{\begin{array}{ll}
    1- \frac{\mu}{r}-\frac{2\Lambda}{d(d-1)} r^2 & (r\ge r_0), \\
    1-\left(\frac{8\pi \rho}{d} +\frac{2\Lambda}{d(d-1)}\right) r^2 & (r<r_0).
    \end{array}\right .
\end{align}
Here the cosmological constant is given by
$\Lambda=- d(d-1)/(2 R_\text{AdS}^2)<0$ with $R_\text{AdS}=1$ and the vacuum energy of the dS region is denoted by $\rho$ that is constant. 
Requiring that the metric function is continuous at the position of the shell, we have
\begin{align}
\label{eq:r0-mu-rho}
    r_0= \left( \frac{d\mu}{8\pi \rho} \right)^{\frac{1}{d}}.
\end{align}

\begin{figure}
\centering
\includegraphics[height=4cm]{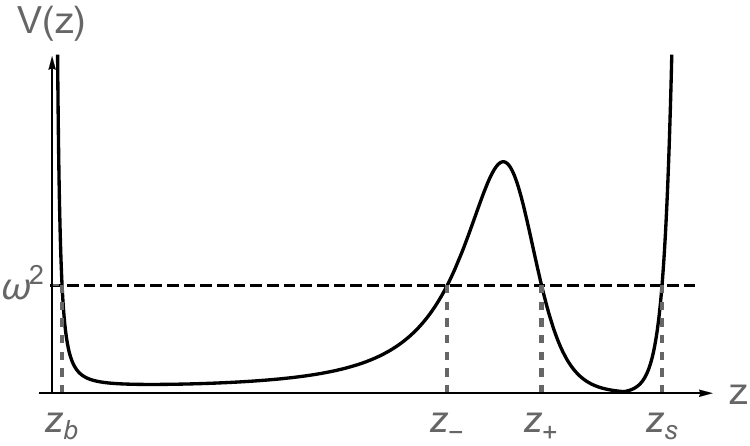}
\caption{The potential for AdS gravastar}
\label{fig:Vgrav}
\end{figure}

We solve the wave equation \eqref{eq:wave} in the WKB approximation also for the AdS gravastar constructed above.
We begin with the case where the equation $\kappa (z) ^2 = 0$ has two zeros at $z_b, z_-$ $(0 \simeq z_b < z_-)$. The form of potential is shown in Fig.\,\ref{fig:Vgrav}.
We first obtain the wave function for $z > z_-$ satisfying the regularity condition at $r \to 0$. We then continue the wave function to the region with $0 \simeq z_b < z < z_-$. From the asymptotic behavior of the wave function, we can obtain the retarded Green function. The expression is the same as in \eqref{eq:gRn}, but now $S(0,z_-)$ is computed with the metric function for AdS gravastar given in \eqref{eq:gravastar-f}.
The bulk-cone singularities correspond to the null geodesics traveling into the gravastar region.

We then move to the case with four zeros in $\kappa (z)^2 = 0$, see Fig.\,\ref{fig:Vgrav}. We set these zeros to be located  at $z= z_b, z_\pm , z_s$ $(0 \simeq z_b < z_- < z_+ < z_s)$. 
We find the wave function for $z > z_s$ satisfying the regularity condition and then continue it to the region with $0 \simeq z_b < z < z_-$. From the asymptotic behavior of the wave function, we can read off the retarded Green function as in \eqref{eq:GRSon}.
Its explicit form is given in \eqref{eq:retardedECO0} with \eqref{eq:retardedECO1} and \eqref{eq:retardedECO2} and it is convenient to expand it in the form of
\begin{align} 
   & G_R (\omega ,\ell) = \frac{\Gamma (- \nu)}{\Gamma(\nu)} 
    \left( \frac{\omega^2 - \ell^2}{4}\right)^\nu \Bigg(\sum_{n=0}^\infty a_n e^{2 i n S(0,z_-)} \nonumber \\
    &+ \sum_{n, p=0}^\infty b_{n,p} e^{2 i n S(0,z_-) + 2 i p S(z_+ , z_s)} e^{-2 S(z_- , z_+)}\Bigg)  \label{eq:GRAB}
\end{align}
up to the terms of order $e^{-4 S(z_- , z_+)}$.
Here we set
\begin{align}
S(z_+,z_s)  =  \int_{z_+}^{z_s} \kappa (z') d z' .
\end{align}
The coefficients $a_n$ are computed as in \eqref{eq:an}
and $b_{n,p}$ are
\begin{align} \label{eq:GRtegravastar2}
\begin{aligned}
& b_{n,0} =  i (-1)^{n} n e^{- i n \pi \nu} \sin ( \pi \nu) , \\
& b_{n,p} =  2 i (-1)^{n + p} n  e^{- i n \pi \nu}  \sin ( \pi \nu)
\end{aligned}
\end{align}
with $p=1,2,3,\ldots$. The contribution from the terms with $a_n$ is the same as that for AdS-Schwarzschild black hole since the region $z < z_-$ is not changed. Thus, these terms lead to bulk-cone singularities associated with null geodesics going around the photon sphere.

The terms with $b_{n,p}$ are suppressed by $e^{- 2 S(z_,z_+)}$ and are regarded as tunneling effects.
We perform the Fourier transformation of the expression in \eqref{eq:GRAB} as
\begin{align}
g_R(t,\theta)  & = \sum_{n=0}^\infty g_R^{(n)} (t , \theta) +  \sum_{n,p=0}^\infty g_R^{(n,p)} (t , \theta) .
\end{align}
Here $g_R^{(n)} (t , \theta)$ are obtained in \eqref{eq:gE2} and  $g_R^{(n,p)} (t , \theta)$ are 
\begin{align} 
& g_R^{(n,p)} (t , \theta) = \tilde b_{n,p} \int_{- \infty + i \delta}^{\infty + i \delta} d \omega  \int _{- \infty + i \delta '}^{\infty + i \delta '}   d k   e^{- 2 S(z_- , z_+) } \nonumber \\
& \quad \times k^{\alpha}(\omega^2 - k^2)^\nu e^{- i \omega t}  e^{2 i n S(0,z_-) + 2 i p S(z_+ , z_s)} 
\label{eq:G2}
\end{align}
with
\begin{align}
\tilde b_{n,p} 
= \frac{\Gamma (-\nu)}{\Gamma (\nu)} 
\frac{e^{- i\frac{ \pi}{2} \alpha + i \pi \alpha \lfloor \frac{\theta}{\pi} \rfloor}}{
2^{\alpha + 1 + 2 \nu}
\pi |\sin \theta|^\alpha \Gamma (1+\alpha) 
}
b_{n,p} .
\end{align}

We evaluate the integrals of \eqref{eq:G2} in the saddle point approximation as above. We set $|S(z_- ,z_+)|$ to be small, where the suppression is not so large.
We look for the saddle points satisfying \eqref{eq:saddlept} but now with
\begin{align}
\Phi (\omega ,k)= - \omega t + k \theta + 2 n S(0,z_-) + 2 p S(z_+ , z_s).
\end{align}
As above, we obtained
\begin{align} \label{eq:gRnt2}
\begin{aligned}
    & g_R^{(n,p)} (t, \theta )  \simeq \tilde b_{n,p}  \frac{(2 n \bar \varrho u_*)^{\alpha + 2 \nu + 3/2}}{(u_*^2 - 1)^{\alpha + \nu + 3/2}} \\
   & \quad  \times\frac{2 \pi i}{\sqrt{ 2 \bar \varrho n (n \frac{d}{du_*}T(u_*) + p \frac{d}{d u_*} \tilde T  (u_*) )}}\\
   & \quad  \times \frac{1}{( n T (u_*) + p \tilde T (u_*) - t + i \epsilon)^{\alpha + 2 \nu + 3/2}} ,
\end{aligned}
\end{align}
where we have defined
\begin{align} 
\label{eq:TThetat}
\begin{aligned}
&\tilde T(u) = 2 u \int_{r(z_s)}^{r(r_+)} \frac{dr}{f(r)} \frac{1}{\sqrt{u^2 - f (r) r^{-2} }} , \\
&\tilde \Theta (u) = 2  \int_{r(z_s)}^{r(z_+)} \frac{dr}{r^2} \frac{1}{\sqrt{u^2 - f (r) r^{-2} }} .
\end{aligned}
\end{align}
These functions correspond, respectively, to the shift of time and angle for one period of null geodesic between $z_+$ and $z_s$. Here $u_*$ is obtained by solving $\theta = n \Theta (u) + p \tilde \Theta (u)  $.
We may treat $S(z_- ,z_+)$ in \eqref{eq:G2} in a similar way as done for $S(0,z_-)$ and $ S(z_+,z_s)$, then the position of singularity is shifted as in \eqref{eq:gRnt2} with real $\epsilon$. The singularity is thus replaced by a bump whose center is at $t = n T (u_*) + p \tilde T(u_*)$. The bumps with $p>0$ are regarded as echoes following bulk-cone singularities associated with null geodesics going around the photon sphere.

\section{Numerical analysis}
\label{sec:numerical}

As in \eqref{eq:GRSon}, the retarded Green function can be obtained from the bulk wave function satisfying \eqref{eq:wave}. We have solved the equation analytically within the WKB approximation. Here we would like to confirm the results numerically.
To this end, we include a smearing factor $\exp(-\frac{\mathrm{Re}(\omega)^2}{\omega_c^2}-\frac{\ell^2}{\ell_c^2})
$ when performing the Fourier transformation \eqref{eq:Fourier}.
The Gaussian factor reduces the singularities to finite bumps in the coordinate basis.
The smearing corresponds to putting some source of wave packets at $(t,\theta)=(0,0)$ on the boundary~\cite{Kinoshita:2023hgc,Terashima:2023mcr}.

\begin{figure*}[htbp]
\centering
\includegraphics[width=17cm]{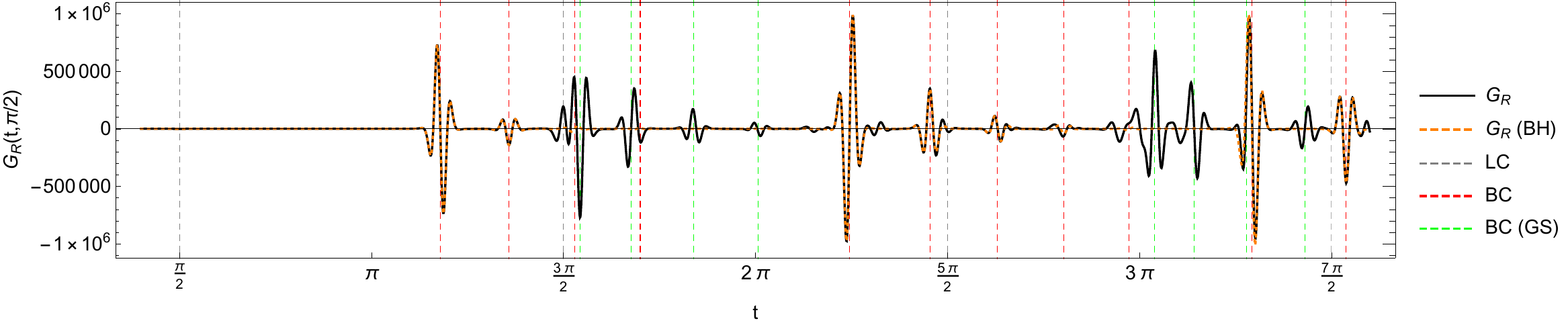}
\caption{The Green function $G_R(t,\pi/2)$ for the gravastar for $d=3$, $\mu=1/15$, and $r_0=1.001r_h$. We adopt the cutoff $\omega_c=\ell_c=35$ and the upper bound $\omega_\mathrm{max}=\ell_\mathrm{max}=150$. The vertical dashed lines indicate the light-cone (LC) and bulk-cone (BC) singularities.}
\label{fig:GRt_GV_3d}
\end{figure*}

A result for AdS gravastar in $d=3$
is shown in Fig.~\ref{fig:GRt_GV_3d}, see \cite{Chen:2025cee,Chen:2025jbf} for other cases.
The black solid and orange dashed curves show $G_R(t,\theta)$ for AdS gravastar and AdS-Schwarzschild black hole with the same $\mu$, respectively.
Some of the bumps are common to the two cases.
They are the wave packets reflected by the photon sphere potential at $r\sim r_c$.
The arrival time expected from the geodesic analysis (red dashed lines) agree with the positions of the bumps.
We can also see the additional bumps specific to the AdS gravastar around $3 \pi /2 \lesssim t \lesssim 2 \pi$ and $3 \pi \lesssim t \lesssim 13 \pi /4$.
They are the waves passing through the gravastar interior.
The geodesic analysis for the arrival time (green dashed lines) is indeed consistent with them.
As expected from the power of the singularities in the semi-analytic formula~\eqref{eq:gRn}, these two kinds of bumps have the same order of magnitude in their amplitude.

\begin{figure}[htbp]
\centering
\includegraphics[width=8cm]{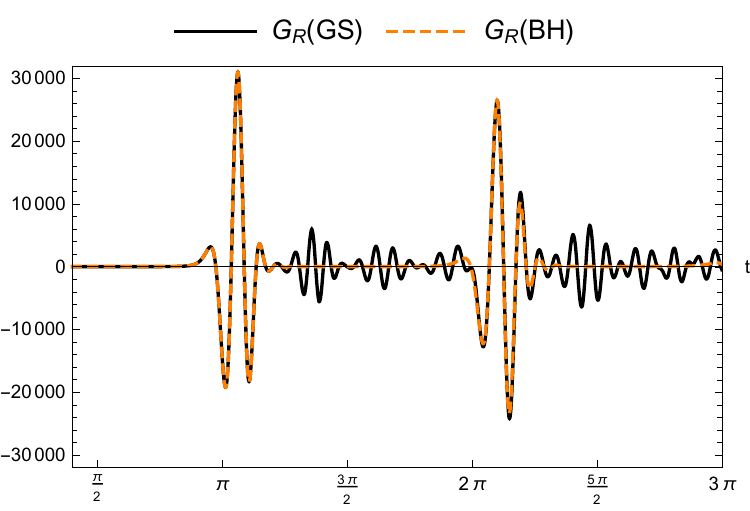}
\caption{Echoes of $G_R(t,\ell)$ for the gravastar with $d=3$, $\mu=1/50$, and $r_0=1.000001r_h$. We take $\ell=1$ and $\omega_c=15$.}
\label{fig:GRtl_GS_d3_mu1_50}
\end{figure}

Along with the bulk-cone singularities, we observe gravitational echoes in the retarded Green function. The Planck constant for the wave equation \eqref{eq:wave} is given by the angular momentum $\ell$ in this case.
In order to clearly see the echoes, we examine the retarded Green function with small fixed $\ell$.
A result is shown in Fig.~\ref{fig:GRtl_GS_d3_mu1_50}. See also \cite{Chen:2025jbf} for more examples.
The black solid and orange dashed curves show $G_R(t,\ell)$ of AdS gravastar and AdS-Schwarzschild black hole with the same $\mu$, respectively.
The common strong bumps are the waves reflected by the photon sphere with $n=1,2$.
The subsequent weak signals are the echoes.

In the astrophysical context, gravitational echoes are also known as a typical signature of ECOs.
The typical time interval for the echoes can be estimated as the propagation time of radial light between the ECO surface and the photon sphere
\begin{align}
    \Delta t_\mathrm{echo}\sim 2\int_{r_0}^{r_c} dr f(r)^{-1} ,
\end{align}
see \cite{Cardoso:2016oxy}.
The interval for the bumps of different $n$ is given by
\begin{align}
    \Delta t_\mathrm{bdry}\sim 2\int_{r_c}^{\infty} dr f(r)^{-1} .
\end{align}
The time width of one wave packet is $\delta t\sim 1/\omega_c$.
In the current case, these time scales, $\delta t\simeq0.067\ll\Delta t_\mathrm{echo}\simeq0.544\ll\Delta t_\mathrm{bdry}\simeq3.243$, are well separated and the echo signals are clearly observed.
The gravitational echoes are observed with a lower cutoff $\omega_c$.
So, they would be the important signature of AdS ECOs when we only have a restricted time resolution for the dual CFTs.

\section{Conclusion and Discussion}
\label{sec:conclusion}

We examined the retarded Green function of a scalar operator from the asymptotically AdS geometry formed by an ECO, i.e., AdS gravastar.
We observed that there are two types of signatures for the horizon-less object, which are bulk-cone singularities and echoes. 
We read off the CFT correlation function from the bulk wave function obtained in the WKB approximation and numerical computations. The analysis in note works for more generic geometry including AdS wormhole as analyzed in \cite{Chen:2025jbf}. 
It is interesting to see what happens if we include charges and rotations to the geometry.

It is desirable to have a direct interpretation from the viewpoint of the dual CFT.
In particular, we can show that the bulk-cone singularities are of the form \eqref{eq:gRn}, where  the power of singularity is generically given by $\alpha + 2 \nu + 3/2 = 2 \Delta - (d-1)/2$ independent of the detailed structure of bulk geometry. 
We would like to understand this universal behavior purely in terms of dual CFT, e.g., by considering the operator product expansion in terms of single and multiple stress tensors.
It is also important to find
the concrete setups of AdS/CFT correspondence and superstring theory.
Furthermore, we would like to examine the gravitational and stringy corrections which may resolve the bulk-cone singularities \cite{Dodelson:2020lal,Dodelson:2023nnr}.
It is expected that the ECOs are unstable due to the localized modes near the minimum of potential and decay into a black hole, see, e.g., \cite{Cunha:2022gde}. The dual CFT description would be useful to understand this issue in particular.

\begin{acknowledgments}
We are grateful to Koji Hashimoto, Yasuyuki Hatsuda, Takaaki Ishii, Keiju Murata, Naritaka Oshita, and Seiji Terashima for useful discussions. The work of H.\,Y.\,C. is supported in part by Ministry of Science and Technology (MOST) through the grant 114-2112-M-002-022-. 
The work of Y.\,H. is supported by JSPS KAKENHI Grant Numbers JP21H05187 and JP23K25867. The work of Y.\,K. is supported by JSPS KAKENHI Grant Number JP23KK0048.
\end{acknowledgments}

\appendix

\section{WKB analysis}
\label{app:WKB}

\subsection{AdS-Schwarzschild black hole}
\label{app:WKB_bh}

We would like to solve the wave equation \eqref{eq:wave} in the WKB approximation for AdS-Schwarzschild black hole. 
We start from the wave function for $z_+ < z  < \infty$. The black hole horizon is located at $z \to \infty$ and the ingoing boundary condition is required at the horizon, i.e.,
 \begin{align}
 \label{eq:bc-ingoing}
     \psi(z) \sim e^{i \omega z}  \quad (z \to \infty) .
 \end{align}
The wave function satisfying the condition is given by
\begin{align} \label{eq:0zzm}
    \psi (z) \sim \frac{1}{\sqrt{\kappa(z)}} e^{i  \int_{z_+}^z d z' \kappa (z')}  .
\end{align}
Applying the WKB connection formula, the  solution for $0 \simeq z_b < z < z_-$ can be found as
\begin{align} \label{eq:psito0}
    \psi (z)\sim \frac{C_+}{\sqrt{\kappa(z)}} e^{i \int_{0}^z d z' \kappa (z')} + \frac{C_-}{\sqrt{\kappa(z)}} e^{ - i \int_{0}^z d z' \kappa (z')} 
\end{align}
with
\begin{align} \label{eq:CmCp}
\begin{aligned}
&C_+ = \left (e^{S(z_- , z_+)} + \frac14e^{-S(z_- , z_+)} \right ) e^{-i S(0,z_-)}, \\ 
&C_- = \left (e^{S(z_- , z_+)} - \frac14e^{-S(z_- , z_+)} \right )e^{ - \frac12 i \pi}  e^{i S(0,z_-)} .
\end{aligned}
\end{align}
For $z < 1/\ell$, the asymptotic geometry is given by the AdS spacetime, and the solution is written in terms of Hankel functions as
\begin{align}
\begin{aligned}
    \psi (z) & =  C_+ \sqrt{\frac{\pi \ell z}{2}} e^{\frac{i \pi \nu}{2} + \frac{i \pi}{4}
     }H^{(1)}_{\nu} (\sqrt{\omega^2-\ell^2} z) \\
     & \quad + C_-  \sqrt{\frac{\pi \ell z}{2}} e^{-\frac{i \pi \nu}{2} - \frac{i \pi}{4} } H^{(2)}_\nu ( \sqrt{\omega^2-\ell^2}z) .
    \end{aligned}
\end{align}
Applying the asymptotic behaviors of Hankel functions,
we can read off the retarded Green function as
\begin{align} \label{eq:retardedAdSG}
\begin{aligned}
   & G_R (\omega ,\ell) =  \frac{\Gamma (- \nu)}{\Gamma(\nu)} \\
    & \quad \times 
    \left( \frac{\omega^2 - \ell^2}{4}\right)^\nu \frac{C_+ e^{-\frac{i \pi \nu}{2} + \frac{i \pi}{4}} - C_- e^{\frac{i \pi \nu}{2} - \frac{i \pi}{4}}}{C_+ e^{\frac{i \pi \nu}{2} + \frac{i \pi}{4}} - C_- e^{-\frac{i \pi \nu}{2} - \frac{i \pi}{4}}}  .
    \end{aligned}
\end{align}
Expanding in $e^{2 i S(0,z_-)}$ and $e^{- 2S(z_-,z_+)}$, we obtain \eqref{eq:GRte} with \eqref{eq:an}.

\subsection{AdS gravastar}
\label{app:WKB_grabastar}

We examine the wave equation \eqref{eq:wave} for the AdS gravastar.
Here we consider the case with four zeros of $\kappa^2 (z) = 0$. We set these zeros to be located  at $z= z_b, z_\pm , z_s$ $(0 \simeq z_b < z_- < z_+ < z_s)$.
The wave function for $z > z_s$ is
\begin{align}
   \psi_{\omega \ell} (z) \sim \frac{1}{\sqrt{q(z)}} e^{ -  \int^z_{z_s} d z' q (z')} .
\end{align}
We connect it to the region $z_+ < z < z_s$ as 
\begin{align}
\begin{aligned}
    \psi (z)&\sim \frac{e^{\frac{i \pi}{4} - i S(z_+,z_s)}}{\sqrt{\kappa(z)}} e^{i  \int^z_{z_+} d z' \kappa (z')}\\
    & \quad + \frac{e^{-\frac{i \pi}{4} + i S(z_+,z_s)}}{\sqrt{\kappa(z)}} e^{- i \int^z_{z_+} d z' \kappa (z')} .
    \end{aligned}
\end{align}
We further continue the wave function to the region $0 \simeq z_b < z_-$ as
\begin{align}
   \psi (z)\sim \frac{ C_+}{\sqrt{\kappa(z)}} e^{i  \int^z_{0} d z' \kappa (z')}  + \frac{ C_-}{\sqrt{ \kappa(z) }} e^{- i  \int^z_{0} d z'  \kappa (z') }
\end{align}
with
\begin{align}
   &  C_+ = e^{\frac{i \pi}{4} - i S(z_+,z_s)} C_+^{(1)} + e^{-\frac{i \pi}{4} + i S(z_+,z_s)} C_-^{(2)} , \\
   & C_- =  e^{\frac{i \pi}{4} - i S(z_+,z_s)} C_- ^{(1)} +  e^{-\frac{i \pi}{4} + i S(z_+,z_s)} C_+ ^{(2)}  .
\end{align}
Here we set
\begin{align}
\begin{aligned}
&C_+ ^{(1)}= \left (e^{S(z_- , z_+)} + \frac14e^{-S(z_- , z_+)} \right ) e^{-i S(0,z_-)}, \\ 
&C_-^{(1)}  = \left (e^{S(z_- , z_+)} - \frac14e^{-S(z_- , z_+)} \right )e^{ - \frac12 i \pi}  e^{i S(0,z_-)} , \\
&C_+ ^{(2)}= \left (e^{S(z_- , z_+)} + \frac14e^{-S(z_- , z_+)} \right ) e^{i S(0,z_-)}, \\ 
&C_-^{(2)}  = \left (e^{S(z_- , z_+)} - \frac14e^{-S(z_- , z_+)} \right )e^{  \frac12 i \pi}  e^{-i S(0,z_-)} .
\end{aligned}
\end{align}
The retarded Green function is given by \eqref{eq:retardedAdSG} for the wave function with the asymptotic behavior \eqref{eq:psito0}.
We thus find
\begin{align} \label{eq:retardedECO0}
    G_R (\omega ,\ell) =  \frac{\Gamma (- \nu)}{\Gamma(\nu)} 
    \left( \frac{\omega^2 - \ell^2}{4}\right)^\nu \frac{P}{Q} 
\end{align}
with
\begin{align}\label{eq:retardedECO1}
&P =  \left(1 + e^{ 2 i S(z_+ ,z_s)}\right) \cos \left( S(0,z_-)+ \frac{ \pi \nu}{2}  \right) \\ 
& - \frac{i}{4} e^{- 2 S(z_- ,z_+)} \left(1 - e^{ 2 i S(z_+ ,z_s)} \right)\sin \left ( S(0,z_-)+ \frac{ \pi \nu}{2}  \right) \nonumber
\end{align}
and
\begin{align}\label{eq:retardedECO2}
&Q = \left(1 + e^{ 2 i S(z_+ ,z_s)} \right)  \cos \left(  S(0,z_-) - \frac{ \pi \nu}{2}  \right) \\ 
&- \frac{i}{4} e^{- 2 S(z_- ,z_+)} \left (1 - e^{ 2 i S(z_+ ,z_s)} \right )\sin \left ( S(0,z_-) - \frac{ \pi \nu}{2}  \right) . \nonumber
\end{align}
Expanding in $e^{2i S(0,z_-)}$, $e^{2i S(z_+,z_s)}$  and  $e^{- 2S(z_-,z_+)}$, we obtain \eqref{eq:GRAB} with \eqref{eq:an} and \eqref{eq:GRtegravastar2}.


\begin{thebibliography}{31}%
\makeatletter
\providecommand \@ifxundefined [1]{%
 \@ifx{#1\undefined}
}%
\providecommand \@ifnum [1]{%
 \ifnum #1\expandafter \@firstoftwo
 \else \expandafter \@secondoftwo
 \fi
}%
\providecommand \@ifx [1]{%
 \ifx #1\expandafter \@firstoftwo
 \else \expandafter \@secondoftwo
 \fi
}%
\providecommand \natexlab [1]{#1}%
\providecommand \enquote  [1]{``#1''}%
\providecommand \bibnamefont  [1]{#1}%
\providecommand \bibfnamefont [1]{#1}%
\providecommand \citenamefont [1]{#1}%
\providecommand \href@noop [0]{\@secondoftwo}%
\providecommand \href [0]{\begingroup \@sanitize@url \@href}%
\providecommand \@href[1]{\@@startlink{#1}\@@href}%
\providecommand \@@href[1]{\endgroup#1\@@endlink}%
\providecommand \@sanitize@url [0]{\catcode `\\12\catcode `\$12\catcode
  `\&12\catcode `\#12\catcode `\^12\catcode `\_12\catcode `\%12\relax}%
\providecommand \@@startlink[1]{}%
\providecommand \@@endlink[0]{}%
\providecommand \url  [0]{\begingroup\@sanitize@url \@url }%
\providecommand \@url [1]{\endgroup\@href {#1}{\urlprefix }}%
\providecommand \urlprefix  [0]{URL }%
\providecommand \Eprint [0]{\href }%
\providecommand \doibase [0]{https://doi.org/}%
\providecommand \selectlanguage [0]{\@gobble}%
\providecommand \bibinfo  [0]{\@secondoftwo}%
\providecommand \bibfield  [0]{\@secondoftwo}%
\providecommand \translation [1]{[#1]}%
\providecommand \BibitemOpen [0]{}%
\providecommand \bibitemStop [0]{}%
\providecommand \bibitemNoStop [0]{.\EOS\space}%
\providecommand \EOS [0]{\spacefactor3000\relax}%
\providecommand \BibitemShut  [1]{\csname bibitem#1\endcsname}%
\let\auto@bib@innerbib\@empty
\bibitem [{\citenamefont {Susskind}(1995)}]{Susskind:1994vu}%
  \BibitemOpen
  \bibfield  {author} {\bibinfo {author} {\bibfnamefont {L.}~\bibnamefont
  {Susskind}},\ }\bibfield  {title} {\bibinfo {title} {{The world as a
  hologram}},\ }\href {https://doi.org/10.1063/1.531249} {\bibfield  {journal}
  {\bibinfo  {journal} {J. Math. Phys.}\ }\textbf {\bibinfo {volume} {36}},\
  \bibinfo {pages} {6377} (\bibinfo {year} {1995})},\ \Eprint
  {https://arxiv.org/abs/hep-th/9409089} {arXiv:hep-th/9409089} \BibitemShut
  {NoStop}%
\bibitem [{\citenamefont {Maldacena}(1998)}]{Maldacena:1997re}%
  \BibitemOpen
  \bibfield  {author} {\bibinfo {author} {\bibfnamefont {J.~M.}\ \bibnamefont
  {Maldacena}},\ }\bibfield  {title} {\bibinfo {title} {{The large $N$ limit of
  superconformal field theories and supergravity}},\ }\href
  {https://doi.org/10.1023/A:1026654312961} {\bibfield  {journal} {\bibinfo
  {journal} {Adv. Theor. Math. Phys.}\ }\textbf {\bibinfo {volume} {2}},\
  \bibinfo {pages} {231} (\bibinfo {year} {1998})},\ \Eprint
  {https://arxiv.org/abs/hep-th/9711200} {arXiv:hep-th/9711200} \BibitemShut
  {NoStop}%
\bibitem [{\citenamefont {Gubser}\ \emph {et~al.}(1998)\citenamefont {Gubser},
  \citenamefont {Klebanov},\ and\ \citenamefont {Polyakov}}]{Gubser:1998bc}%
  \BibitemOpen
  \bibfield  {author} {\bibinfo {author} {\bibfnamefont {S.~S.}\ \bibnamefont
  {Gubser}}, \bibinfo {author} {\bibfnamefont {I.~R.}\ \bibnamefont
  {Klebanov}},\ and\ \bibinfo {author} {\bibfnamefont {A.~M.}\ \bibnamefont
  {Polyakov}},\ }\bibfield  {title} {\bibinfo {title} {{Gauge theory
  correlators from noncritical string theory}},\ }\href
  {https://doi.org/10.1016/S0370-2693(98)00377-3} {\bibfield  {journal}
  {\bibinfo  {journal} {Phys. Lett. B}\ }\textbf {\bibinfo {volume} {428}},\
  \bibinfo {pages} {105} (\bibinfo {year} {1998})},\ \Eprint
  {https://arxiv.org/abs/hep-th/9802109} {arXiv:hep-th/9802109} \BibitemShut
  {NoStop}%
\bibitem [{\citenamefont {Witten}(1998)}]{Witten:1998qj}%
  \BibitemOpen
  \bibfield  {author} {\bibinfo {author} {\bibfnamefont {E.}~\bibnamefont
  {Witten}},\ }\bibfield  {title} {\bibinfo {title} {{Anti-de Sitter space and
  holography}},\ }\href {https://doi.org/10.4310/ATMP.1998.v2.n2.a2} {\bibfield
   {journal} {\bibinfo  {journal} {Adv. Theor. Math. Phys.}\ }\textbf {\bibinfo
  {volume} {2}},\ \bibinfo {pages} {253} (\bibinfo {year} {1998})},\ \Eprint
  {https://arxiv.org/abs/hep-th/9802150} {arXiv:hep-th/9802150} \BibitemShut
  {NoStop}%
\bibitem [{\citenamefont {Almheiri}\ \emph {et~al.}(2013)\citenamefont
  {Almheiri}, \citenamefont {Marolf}, \citenamefont {Polchinski},\ and\
  \citenamefont {Sully}}]{Almheiri:2012rt}%
  \BibitemOpen
  \bibfield  {author} {\bibinfo {author} {\bibfnamefont {A.}~\bibnamefont
  {Almheiri}}, \bibinfo {author} {\bibfnamefont {D.}~\bibnamefont {Marolf}},
  \bibinfo {author} {\bibfnamefont {J.}~\bibnamefont {Polchinski}},\ and\
  \bibinfo {author} {\bibfnamefont {J.}~\bibnamefont {Sully}},\ }\bibfield
  {title} {\bibinfo {title} {{Black holes: Complementarity or firewalls?}},\
  }\href {https://doi.org/10.1007/JHEP02(2013)062} {\bibfield  {journal}
  {\bibinfo  {journal} {JHEP}\ }\textbf {\bibinfo {volume} {02}},\ \bibinfo
  {pages} {062}},\ \Eprint {https://arxiv.org/abs/1207.3123} {arXiv:1207.3123
  [hep-th]} \BibitemShut {NoStop}%
\bibitem [{\citenamefont {Lunin}\ and\ \citenamefont
  {Mathur}(2002{\natexlab{a}})}]{Lunin:2001jy}%
  \BibitemOpen
  \bibfield  {author} {\bibinfo {author} {\bibfnamefont {O.}~\bibnamefont
  {Lunin}}\ and\ \bibinfo {author} {\bibfnamefont {S.~D.}\ \bibnamefont
  {Mathur}},\ }\bibfield  {title} {\bibinfo {title} {{AdS/CFT duality and the
  black hole information paradox}},\ }\href
  {https://doi.org/10.1016/S0550-3213(01)00620-4} {\bibfield  {journal}
  {\bibinfo  {journal} {Nucl. Phys. B}\ }\textbf {\bibinfo {volume} {623}},\
  \bibinfo {pages} {342} (\bibinfo {year} {2002}{\natexlab{a}})},\ \Eprint
  {https://arxiv.org/abs/hep-th/0109154} {arXiv:hep-th/0109154} \BibitemShut
  {NoStop}%
\bibitem [{\citenamefont {Lunin}\ and\ \citenamefont
  {Mathur}(2002{\natexlab{b}})}]{Lunin:2002qf}%
  \BibitemOpen
  \bibfield  {author} {\bibinfo {author} {\bibfnamefont {O.}~\bibnamefont
  {Lunin}}\ and\ \bibinfo {author} {\bibfnamefont {S.~D.}\ \bibnamefont
  {Mathur}},\ }\bibfield  {title} {\bibinfo {title} {{Statistical
  interpretation of Bekenstein entropy for systems with a stretched horizon}},\
  }\href {https://doi.org/10.1103/PhysRevLett.88.211303} {\bibfield  {journal}
  {\bibinfo  {journal} {Phys. Rev. Lett.}\ }\textbf {\bibinfo {volume} {88}},\
  \bibinfo {pages} {211303} (\bibinfo {year} {2002}{\natexlab{b}})},\ \Eprint
  {https://arxiv.org/abs/hep-th/0202072} {arXiv:hep-th/0202072} \BibitemShut
  {NoStop}%
\bibitem [{\citenamefont {Cardoso}\ and\ \citenamefont
  {Pani}(2019)}]{Cardoso:2019rvt}%
  \BibitemOpen
  \bibfield  {author} {\bibinfo {author} {\bibfnamefont {V.}~\bibnamefont
  {Cardoso}}\ and\ \bibinfo {author} {\bibfnamefont {P.}~\bibnamefont {Pani}},\
  }\bibfield  {title} {\bibinfo {title} {{Testing the nature of dark compact
  objects: A status report}},\ }\href
  {https://doi.org/10.1007/s41114-019-0020-4} {\bibfield  {journal} {\bibinfo
  {journal} {Living Rev. Rel.}\ }\textbf {\bibinfo {volume} {22}},\ \bibinfo
  {pages} {4} (\bibinfo {year} {2019})},\ \Eprint
  {https://arxiv.org/abs/1904.05363} {arXiv:1904.05363 [gr-qc]} \BibitemShut
  {NoStop}%
\bibitem [{\citenamefont {Mazur}\ and\ \citenamefont
  {Mottola}(2023)}]{Mazur:2001fv}%
  \BibitemOpen
  \bibfield  {author} {\bibinfo {author} {\bibfnamefont {P.~O.}\ \bibnamefont
  {Mazur}}\ and\ \bibinfo {author} {\bibfnamefont {E.}~\bibnamefont
  {Mottola}},\ }\bibfield  {title} {\bibinfo {title} {{Gravitational condensate
  stars: An alternative to black holes}},\ }\href
  {https://doi.org/10.3390/universe9020088} {\bibfield  {journal} {\bibinfo
  {journal} {Universe}\ }\textbf {\bibinfo {volume} {9}},\ \bibinfo {pages}
  {88} (\bibinfo {year} {2023})},\ \Eprint
  {https://arxiv.org/abs/gr-qc/0109035} {arXiv:gr-qc/0109035} \BibitemShut
  {NoStop}%
\bibitem [{\citenamefont {Visser}\ and\ \citenamefont
  {Wiltshire}(2004)}]{Visser:2003ge}%
  \BibitemOpen
  \bibfield  {author} {\bibinfo {author} {\bibfnamefont {M.}~\bibnamefont
  {Visser}}\ and\ \bibinfo {author} {\bibfnamefont {D.~L.}\ \bibnamefont
  {Wiltshire}},\ }\bibfield  {title} {\bibinfo {title} {{Stable gravastars: An
  alternative to black holes?}},\ }\href
  {https://doi.org/10.1088/0264-9381/21/4/027} {\bibfield  {journal} {\bibinfo
  {journal} {Class. Quant. Grav.}\ }\textbf {\bibinfo {volume} {21}},\ \bibinfo
  {pages} {1135} (\bibinfo {year} {2004})},\ \Eprint
  {https://arxiv.org/abs/gr-qc/0310107} {arXiv:gr-qc/0310107} \BibitemShut
  {NoStop}%
\bibitem [{\citenamefont {Pani}\ \emph {et~al.}(2009)\citenamefont {Pani},
  \citenamefont {Berti}, \citenamefont {Cardoso}, \citenamefont {Chen},\ and\
  \citenamefont {Norte}}]{Pani:2009ss}%
  \BibitemOpen
  \bibfield  {author} {\bibinfo {author} {\bibfnamefont {P.}~\bibnamefont
  {Pani}}, \bibinfo {author} {\bibfnamefont {E.}~\bibnamefont {Berti}},
  \bibinfo {author} {\bibfnamefont {V.}~\bibnamefont {Cardoso}}, \bibinfo
  {author} {\bibfnamefont {Y.}~\bibnamefont {Chen}},\ and\ \bibinfo {author}
  {\bibfnamefont {R.}~\bibnamefont {Norte}},\ }\bibfield  {title} {\bibinfo
  {title} {{Gravitational wave signatures of the absence of an event horizon.
  I. Nonradial oscillations of a thin-shell gravastar}},\ }\href
  {https://doi.org/10.1103/PhysRevD.80.124047} {\bibfield  {journal} {\bibinfo
  {journal} {Phys. Rev. D}\ }\textbf {\bibinfo {volume} {80}},\ \bibinfo
  {pages} {124047} (\bibinfo {year} {2009})},\ \Eprint
  {https://arxiv.org/abs/0909.0287} {arXiv:0909.0287 [gr-qc]} \BibitemShut
  {NoStop}%
\bibitem [{\citenamefont {Cardoso}\ \emph {et~al.}(2014)\citenamefont
  {Cardoso}, \citenamefont {Crispino}, \citenamefont {Macedo}, \citenamefont
  {Okawa},\ and\ \citenamefont {Pani}}]{Cardoso:2014sna}%
  \BibitemOpen
  \bibfield  {author} {\bibinfo {author} {\bibfnamefont {V.}~\bibnamefont
  {Cardoso}}, \bibinfo {author} {\bibfnamefont {L.~C.~B.}\ \bibnamefont
  {Crispino}}, \bibinfo {author} {\bibfnamefont {C.~F.~B.}\ \bibnamefont
  {Macedo}}, \bibinfo {author} {\bibfnamefont {H.}~\bibnamefont {Okawa}},\ and\
  \bibinfo {author} {\bibfnamefont {P.}~\bibnamefont {Pani}},\ }\bibfield
  {title} {\bibinfo {title} {{Light rings as observational evidence for event
  horizons: Long-lived modes, ergoregions and nonlinear instabilities of
  ultracompact objects}},\ }\href {https://doi.org/10.1103/PhysRevD.90.044069}
  {\bibfield  {journal} {\bibinfo  {journal} {Phys. Rev. D}\ }\textbf {\bibinfo
  {volume} {90}},\ \bibinfo {pages} {044069} (\bibinfo {year} {2014})},\
  \Eprint {https://arxiv.org/abs/1406.5510} {arXiv:1406.5510 [gr-qc]}
  \BibitemShut {NoStop}%
\bibitem [{\citenamefont {Chen}\ \emph {et~al.}(2025)\citenamefont {Chen},
  \citenamefont {Hikida},\ and\ \citenamefont {Koga}}]{Chen:2025cee}%
  \BibitemOpen
  \bibfield  {author} {\bibinfo {author} {\bibfnamefont {H.-Y.}\ \bibnamefont
  {Chen}}, \bibinfo {author} {\bibfnamefont {Y.}~\bibnamefont {Hikida}},\ and\
  \bibinfo {author} {\bibfnamefont {Y.}~\bibnamefont {Koga}},\ }\bibfield
  {title} {\bibinfo {title} {{AdS gravastar and bulk-cone singularities}},\
  }\href {https://doi.org/10.1007/JHEP07(2025)199} {\bibfield  {journal}
  {\bibinfo  {journal} {JHEP}\ }\textbf {\bibinfo {volume} {07}},\ \bibinfo
  {pages} {199}},\ \Eprint {https://arxiv.org/abs/2502.11403} {arXiv:2502.11403
  [hep-th]} \BibitemShut {NoStop}%
\bibitem [{\citenamefont {Chen}\ \emph {et~al.}(2026)\citenamefont {Chen},
  \citenamefont {Hikida},\ and\ \citenamefont {Koga}}]{Chen:2025jbf}%
  \BibitemOpen
  \bibfield  {author} {\bibinfo {author} {\bibfnamefont {H.-Y.}\ \bibnamefont
  {Chen}}, \bibinfo {author} {\bibfnamefont {Y.}~\bibnamefont {Hikida}},\ and\
  \bibinfo {author} {\bibfnamefont {Y.}~\bibnamefont {Koga}},\ }\bibfield
  {title} {\bibinfo {title} {{Bulk-cone singularities and echos from AdS exotic
  compact objects}},\ }\href {https://doi.org/10.1007/JHEP04(2026)210}
  {\bibfield  {journal} {\bibinfo  {journal} {JHEP}\ }\textbf {\bibinfo
  {volume} {04}},\ \bibinfo {pages} {210}},\ \Eprint
  {https://arxiv.org/abs/2512.21535} {arXiv:2512.21535 [hep-th]} \BibitemShut
  {NoStop}%
\bibitem [{\citenamefont {Son}\ and\ \citenamefont
  {Starinets}(2002)}]{Son:2002sd}%
  \BibitemOpen
  \bibfield  {author} {\bibinfo {author} {\bibfnamefont {D.~T.}\ \bibnamefont
  {Son}}\ and\ \bibinfo {author} {\bibfnamefont {A.~O.}\ \bibnamefont
  {Starinets}},\ }\bibfield  {title} {\bibinfo {title} {{Minkowski space
  correlators in AdS/CFT correspondence: Recipe and applications}},\ }\href
  {https://doi.org/10.1088/1126-6708/2002/09/042} {\bibfield  {journal}
  {\bibinfo  {journal} {JHEP}\ }\textbf {\bibinfo {volume} {09}},\ \bibinfo
  {pages} {042}},\ \Eprint {https://arxiv.org/abs/hep-th/0205051}
  {arXiv:hep-th/0205051} \BibitemShut {NoStop}%
\bibitem [{\citenamefont {Hubeny}\ \emph {et~al.}(2007)\citenamefont {Hubeny},
  \citenamefont {Liu},\ and\ \citenamefont {Rangamani}}]{Hubeny:2006yu}%
  \BibitemOpen
  \bibfield  {author} {\bibinfo {author} {\bibfnamefont {V.~E.}\ \bibnamefont
  {Hubeny}}, \bibinfo {author} {\bibfnamefont {H.}~\bibnamefont {Liu}},\ and\
  \bibinfo {author} {\bibfnamefont {M.}~\bibnamefont {Rangamani}},\ }\bibfield
  {title} {\bibinfo {title} {{Bulk-cone singularities \& signatures of horizon
  formation in AdS/CFT}},\ }\href
  {https://doi.org/10.1088/1126-6708/2007/01/009} {\bibfield  {journal}
  {\bibinfo  {journal} {JHEP}\ }\textbf {\bibinfo {volume} {01}},\ \bibinfo
  {pages} {009}},\ \Eprint {https://arxiv.org/abs/hep-th/0610041}
  {arXiv:hep-th/0610041} \BibitemShut {NoStop}%
\bibitem [{\citenamefont {Maldacena}\ \emph {et~al.}(2017)\citenamefont
  {Maldacena}, \citenamefont {Simmons-Duffin},\ and\ \citenamefont
  {Zhiboedov}}]{Maldacena:2015iua}%
  \BibitemOpen
  \bibfield  {author} {\bibinfo {author} {\bibfnamefont {J.}~\bibnamefont
  {Maldacena}}, \bibinfo {author} {\bibfnamefont {D.}~\bibnamefont
  {Simmons-Duffin}},\ and\ \bibinfo {author} {\bibfnamefont {A.}~\bibnamefont
  {Zhiboedov}},\ }\bibfield  {title} {\bibinfo {title} {{Looking for a bulk
  point}},\ }\href {https://doi.org/10.1007/JHEP01(2017)013} {\bibfield
  {journal} {\bibinfo  {journal} {JHEP}\ }\textbf {\bibinfo {volume} {01}},\
  \bibinfo {pages} {013}},\ \Eprint {https://arxiv.org/abs/1509.03612}
  {arXiv:1509.03612 [hep-th]} \BibitemShut {NoStop}%
\bibitem [{\citenamefont {Hashimoto}\ \emph {et~al.}(2020)\citenamefont
  {Hashimoto}, \citenamefont {Kinoshita},\ and\ \citenamefont
  {Murata}}]{Hashimoto:2018okj}%
  \BibitemOpen
  \bibfield  {author} {\bibinfo {author} {\bibfnamefont {K.}~\bibnamefont
  {Hashimoto}}, \bibinfo {author} {\bibfnamefont {S.}~\bibnamefont
  {Kinoshita}},\ and\ \bibinfo {author} {\bibfnamefont {K.}~\bibnamefont
  {Murata}},\ }\bibfield  {title} {\bibinfo {title} {{Imaging black holes
  through the AdS/CFT correspondence}},\ }\href
  {https://doi.org/10.1103/PhysRevD.101.066018} {\bibfield  {journal} {\bibinfo
   {journal} {Phys. Rev. D}\ }\textbf {\bibinfo {volume} {101}},\ \bibinfo
  {pages} {066018} (\bibinfo {year} {2020})},\ \Eprint
  {https://arxiv.org/abs/1811.12617} {arXiv:1811.12617 [hep-th]} \BibitemShut
  {NoStop}%
\bibitem [{\citenamefont {Hashimoto}\ \emph {et~al.}(2019)\citenamefont
  {Hashimoto}, \citenamefont {Kinoshita},\ and\ \citenamefont
  {Murata}}]{Hashimoto:2019jmw}%
  \BibitemOpen
  \bibfield  {author} {\bibinfo {author} {\bibfnamefont {K.}~\bibnamefont
  {Hashimoto}}, \bibinfo {author} {\bibfnamefont {S.}~\bibnamefont
  {Kinoshita}},\ and\ \bibinfo {author} {\bibfnamefont {K.}~\bibnamefont
  {Murata}},\ }\bibfield  {title} {\bibinfo {title} {{Einstein rings in
  holography}},\ }\href {https://doi.org/10.1103/PhysRevLett.123.031602}
  {\bibfield  {journal} {\bibinfo  {journal} {Phys. Rev. Lett.}\ }\textbf
  {\bibinfo {volume} {123}},\ \bibinfo {pages} {031602} (\bibinfo {year}
  {2019})},\ \Eprint {https://arxiv.org/abs/1906.09113} {arXiv:1906.09113
  [hep-th]} \BibitemShut {NoStop}%
\bibitem [{\citenamefont {Dodelson}\ and\ \citenamefont
  {Ooguri}(2021)}]{Dodelson:2020lal}%
  \BibitemOpen
  \bibfield  {author} {\bibinfo {author} {\bibfnamefont {M.}~\bibnamefont
  {Dodelson}}\ and\ \bibinfo {author} {\bibfnamefont {H.}~\bibnamefont
  {Ooguri}},\ }\bibfield  {title} {\bibinfo {title} {{Singularities of thermal
  correlators at strong coupling}},\ }\href
  {https://doi.org/10.1103/PhysRevD.103.066018} {\bibfield  {journal} {\bibinfo
   {journal} {Phys. Rev. D}\ }\textbf {\bibinfo {volume} {103}},\ \bibinfo
  {pages} {066018} (\bibinfo {year} {2021})},\ \Eprint
  {https://arxiv.org/abs/2010.09734} {arXiv:2010.09734 [hep-th]} \BibitemShut
  {NoStop}%
\bibitem [{\citenamefont {Kinoshita}\ \emph {et~al.}(2023)\citenamefont
  {Kinoshita}, \citenamefont {Murata},\ and\ \citenamefont
  {Takeda}}]{Kinoshita:2023hgc}%
  \BibitemOpen
  \bibfield  {author} {\bibinfo {author} {\bibfnamefont {S.}~\bibnamefont
  {Kinoshita}}, \bibinfo {author} {\bibfnamefont {K.}~\bibnamefont {Murata}},\
  and\ \bibinfo {author} {\bibfnamefont {D.}~\bibnamefont {Takeda}},\
  }\bibfield  {title} {\bibinfo {title} {{Shooting null geodesics into
  holographic spacetimes}},\ }\href {https://doi.org/10.1007/JHEP10(2023)074}
  {\bibfield  {journal} {\bibinfo  {journal} {JHEP}\ }\textbf {\bibinfo
  {volume} {10}},\ \bibinfo {pages} {074}},\ \Eprint
  {https://arxiv.org/abs/2304.01936} {arXiv:2304.01936 [hep-th]} \BibitemShut
  {NoStop}%
\bibitem [{\citenamefont {Terashima}(2024)}]{Terashima:2023mcr}%
  \BibitemOpen
  \bibfield  {author} {\bibinfo {author} {\bibfnamefont {S.}~\bibnamefont
  {Terashima}},\ }\bibfield  {title} {\bibinfo {title} {{Wave packets in
  AdS/CFT correspondence}},\ }\href
  {https://doi.org/10.1103/PhysRevD.109.106012} {\bibfield  {journal} {\bibinfo
   {journal} {Phys. Rev. D}\ }\textbf {\bibinfo {volume} {109}},\ \bibinfo
  {pages} {106012} (\bibinfo {year} {2024})},\ \Eprint
  {https://arxiv.org/abs/2304.08478} {arXiv:2304.08478 [hep-th]} \BibitemShut
  {NoStop}%
\bibitem [{\citenamefont {Hashimoto}\ \emph {et~al.}(2023)\citenamefont
  {Hashimoto}, \citenamefont {Sugiura}, \citenamefont {Sugiyama},\ and\
  \citenamefont {Yoda}}]{Hashimoto:2023buz}%
  \BibitemOpen
  \bibfield  {author} {\bibinfo {author} {\bibfnamefont {K.}~\bibnamefont
  {Hashimoto}}, \bibinfo {author} {\bibfnamefont {K.}~\bibnamefont {Sugiura}},
  \bibinfo {author} {\bibfnamefont {K.}~\bibnamefont {Sugiyama}},\ and\
  \bibinfo {author} {\bibfnamefont {T.}~\bibnamefont {Yoda}},\ }\bibfield
  {title} {\bibinfo {title} {{Photon sphere and quasinormal modes in
  AdS/CFT}},\ }\href {https://doi.org/10.1007/JHEP10(2023)149} {\bibfield
  {journal} {\bibinfo  {journal} {JHEP}\ }\textbf {\bibinfo {volume} {10}},\
  \bibinfo {pages} {149}},\ \Eprint {https://arxiv.org/abs/2307.00237}
  {arXiv:2307.00237 [hep-th]} \BibitemShut {NoStop}%
\bibitem [{\citenamefont {Dodelson}\ \emph {et~al.}(2024)\citenamefont
  {Dodelson}, \citenamefont {Iossa}, \citenamefont {Karlsson}, \citenamefont
  {Lupsasca},\ and\ \citenamefont {Zhiboedov}}]{Dodelson:2023nnr}%
  \BibitemOpen
  \bibfield  {author} {\bibinfo {author} {\bibfnamefont {M.}~\bibnamefont
  {Dodelson}}, \bibinfo {author} {\bibfnamefont {C.}~\bibnamefont {Iossa}},
  \bibinfo {author} {\bibfnamefont {R.}~\bibnamefont {Karlsson}}, \bibinfo
  {author} {\bibfnamefont {A.}~\bibnamefont {Lupsasca}},\ and\ \bibinfo
  {author} {\bibfnamefont {A.}~\bibnamefont {Zhiboedov}},\ }\bibfield  {title}
  {\bibinfo {title} {{Black hole bulk-cone singularities}},\ }\href
  {https://doi.org/10.1007/JHEP07(2024)046} {\bibfield  {journal} {\bibinfo
  {journal} {JHEP}\ }\textbf {\bibinfo {volume} {07}},\ \bibinfo {pages}
  {046}},\ \Eprint {https://arxiv.org/abs/2310.15236} {arXiv:2310.15236
  [hep-th]} \BibitemShut {NoStop}%
\bibitem [{\citenamefont {Caron-Huot}\ \emph {et~al.}(2025)\citenamefont
  {Caron-Huot}, \citenamefont {Chakravarty},\ and\ \citenamefont
  {Namjou}}]{Caron-Huot:2025she}%
  \BibitemOpen
  \bibfield  {author} {\bibinfo {author} {\bibfnamefont {S.}~\bibnamefont
  {Caron-Huot}}, \bibinfo {author} {\bibfnamefont {J.}~\bibnamefont
  {Chakravarty}},\ and\ \bibinfo {author} {\bibfnamefont {K.}~\bibnamefont
  {Namjou}},\ }\bibfield  {title} {\bibinfo {title} {{Boundary imprint of bulk
  causality}},\ }\href {https://doi.org/10.1007/JHEP07(2025)076} {\bibfield
  {journal} {\bibinfo  {journal} {JHEP}\ }\textbf {\bibinfo {volume} {07}},\
  \bibinfo {pages} {076}},\ \Eprint {https://arxiv.org/abs/2501.13182}
  {arXiv:2501.13182 [hep-th]} \BibitemShut {NoStop}%
\bibitem [{\citenamefont {Cardoso}\ \emph
  {et~al.}(2016{\natexlab{a}})\citenamefont {Cardoso}, \citenamefont
  {Franzin},\ and\ \citenamefont {Pani}}]{Cardoso:2016rao}%
  \BibitemOpen
  \bibfield  {author} {\bibinfo {author} {\bibfnamefont {V.}~\bibnamefont
  {Cardoso}}, \bibinfo {author} {\bibfnamefont {E.}~\bibnamefont {Franzin}},\
  and\ \bibinfo {author} {\bibfnamefont {P.}~\bibnamefont {Pani}},\ }\bibfield
  {title} {\bibinfo {title} {{Is the gravitational-wave ringdown a probe of the
  event horizon?}},\ }\href {https://doi.org/10.1103/PhysRevLett.116.171101}
  {\bibfield  {journal} {\bibinfo  {journal} {Phys. Rev. Lett.}\ }\textbf
  {\bibinfo {volume} {116}},\ \bibinfo {pages} {171101} (\bibinfo {year}
  {2016}{\natexlab{a}})},\ \bibinfo {note} {[Erratum: Phys.Rev.Lett. 117,
  089902 (2016)]},\ \Eprint {https://arxiv.org/abs/1602.07309}
  {arXiv:1602.07309 [gr-qc]} \BibitemShut {NoStop}%
\bibitem [{\citenamefont {Cardoso}\ \emph
  {et~al.}(2016{\natexlab{b}})\citenamefont {Cardoso}, \citenamefont {Hopper},
  \citenamefont {Macedo}, \citenamefont {Palenzuela},\ and\ \citenamefont
  {Pani}}]{Cardoso:2016oxy}%
  \BibitemOpen
  \bibfield  {author} {\bibinfo {author} {\bibfnamefont {V.}~\bibnamefont
  {Cardoso}}, \bibinfo {author} {\bibfnamefont {S.}~\bibnamefont {Hopper}},
  \bibinfo {author} {\bibfnamefont {C.~F.~B.}\ \bibnamefont {Macedo}}, \bibinfo
  {author} {\bibfnamefont {C.}~\bibnamefont {Palenzuela}},\ and\ \bibinfo
  {author} {\bibfnamefont {P.}~\bibnamefont {Pani}},\ }\bibfield  {title}
  {\bibinfo {title} {{Gravitational-wave signatures of exotic compact objects
  and of quantum corrections at the horizon scale}},\ }\href
  {https://doi.org/10.1103/PhysRevD.94.084031} {\bibfield  {journal} {\bibinfo
  {journal} {Phys. Rev. D}\ }\textbf {\bibinfo {volume} {94}},\ \bibinfo
  {pages} {084031} (\bibinfo {year} {2016}{\natexlab{b}})},\ \Eprint
  {https://arxiv.org/abs/1608.08637} {arXiv:1608.08637 [gr-qc]} \BibitemShut
  {NoStop}%
\bibitem [{\citenamefont {Oshita}\ and\ \citenamefont
  {Afshordi}(2019)}]{Oshita:2018fqu}%
  \BibitemOpen
  \bibfield  {author} {\bibinfo {author} {\bibfnamefont {N.}~\bibnamefont
  {Oshita}}\ and\ \bibinfo {author} {\bibfnamefont {N.}~\bibnamefont
  {Afshordi}},\ }\bibfield  {title} {\bibinfo {title} {{Probing microstructure
  of black hole spacetimes with gravitational wave echoes}},\ }\href
  {https://doi.org/10.1103/PhysRevD.99.044002} {\bibfield  {journal} {\bibinfo
  {journal} {Phys. Rev. D}\ }\textbf {\bibinfo {volume} {99}},\ \bibinfo
  {pages} {044002} (\bibinfo {year} {2019})},\ \Eprint
  {https://arxiv.org/abs/1807.10287} {arXiv:1807.10287 [gr-qc]} \BibitemShut
  {NoStop}%
\bibitem [{\citenamefont {Oshita}\ \emph {et~al.}(2020)\citenamefont {Oshita},
  \citenamefont {Tsuna},\ and\ \citenamefont {Afshordi}}]{Oshita:2020dox}%
  \BibitemOpen
  \bibfield  {author} {\bibinfo {author} {\bibfnamefont {N.}~\bibnamefont
  {Oshita}}, \bibinfo {author} {\bibfnamefont {D.}~\bibnamefont {Tsuna}},\ and\
  \bibinfo {author} {\bibfnamefont {N.}~\bibnamefont {Afshordi}},\ }\bibfield
  {title} {\bibinfo {title} {{Quantum black hole seismology I: Echoes,
  ergospheres, and spectra}},\ }\href
  {https://doi.org/10.1103/PhysRevD.102.024045} {\bibfield  {journal} {\bibinfo
   {journal} {Phys. Rev. D}\ }\textbf {\bibinfo {volume} {102}},\ \bibinfo
  {pages} {024045} (\bibinfo {year} {2020})},\ \Eprint
  {https://arxiv.org/abs/2001.11642} {arXiv:2001.11642 [gr-qc]} \BibitemShut
  {NoStop}%
\bibitem [{\citenamefont {Terashima}(2025)}]{Terashima:2025tct}%
  \BibitemOpen
  \bibfield  {author} {\bibinfo {author} {\bibfnamefont {S.}~\bibnamefont
  {Terashima}},\ }\bibfield  {title} {\bibinfo {title} {{Stretched horizon
  dissipation and the fate of echoes}},\ }\href
  {https://doi.org/10.1007/JHEP10(2025)147} {\bibfield  {journal} {\bibinfo
  {journal} {JHEP}\ }\textbf {\bibinfo {volume} {10}},\ \bibinfo {pages}
  {147}},\ \Eprint {https://arxiv.org/abs/2506.20462} {arXiv:2506.20462
  [hep-th]} \BibitemShut {NoStop}%
\bibitem [{\citenamefont {Cunha}\ \emph {et~al.}(2023)\citenamefont {Cunha},
  \citenamefont {Herdeiro}, \citenamefont {Radu},\ and\ \citenamefont
  {Sanchis-Gual}}]{Cunha:2022gde}%
  \BibitemOpen
  \bibfield  {author} {\bibinfo {author} {\bibfnamefont {P.~V.~P.}\
  \bibnamefont {Cunha}}, \bibinfo {author} {\bibfnamefont {C.}~\bibnamefont
  {Herdeiro}}, \bibinfo {author} {\bibfnamefont {E.}~\bibnamefont {Radu}},\
  and\ \bibinfo {author} {\bibfnamefont {N.}~\bibnamefont {Sanchis-Gual}},\
  }\bibfield  {title} {\bibinfo {title} {{Exotic compact objects and the fate
  of the light-ring instability}},\ }\href
  {https://doi.org/10.1103/PhysRevLett.130.061401} {\bibfield  {journal}
  {\bibinfo  {journal} {Phys. Rev. Lett.}\ }\textbf {\bibinfo {volume} {130}},\
  \bibinfo {pages} {061401} (\bibinfo {year} {2023})},\ \Eprint
  {https://arxiv.org/abs/2207.13713} {arXiv:2207.13713 [gr-qc]} \BibitemShut
  {NoStop}%
\end{thebibliography}

%

\end{document}